\documentclass[%
reprint,
amsmath,amssymb,
aps,
]{revtex4-2}
\usepackage{xcolor}
\usepackage{graphicx}
\usepackage{dcolumn}
\graphicspath{{./figures/}}
\usepackage{enumerate}
\usepackage{subcaption}
\usepackage{bm}

\begin{document}
	
\preprint{APS/123-QED}

\title{Dynamic shapes of floppy vesicles enclosing active Brownian particles with membrane adhesion}

\author{Priyanka Iyer, Gerhard Gompper, and Dmitry A. Fedosov}
\altaffiliation{Theoretical Physics of Living Matter, Institute of Biological Information Processing and Institute for Advanced Simulation, Forschungszentrum J\"ulich, 52425 J\"ulich, Germany \\ Email: p.iyer@fz-juelich.de, g.gompper@fz-juelich.de, d.fedosov@fz-juelich.de}

\date{\today}

\begin{abstract}
 Recent advances in micro- and nano-technologies allow the construction of complex active systems from biological and synthetic materials. An interesting example 
is active vesicles, which consist of a membrane enclosing self-propelled particles, and exhibit several features resembling biological cells.
We investigate numerically the behavior of active vesicles, where the enclosed self-propelled particles can adhere to the membrane. A vesicle is represented 
by a dynamically triangulated membrane, while the adhesive active particles are modelled as active Brownian particles (ABPs) that interact with the membrane 
via the Lennard-Jones potential. Phase diagrams of dynamic vesicle shapes as a function of ABP activity and particle volume fraction inside the vesicle are constructed 
for different strengths of adhesive interactions. At low ABP activity, adhesive interactions dominate over the propulsion forces, such that the vesicle attains near 
static configurations, with protrusions of membrane-wrapped ABPs having ring-like and sheet-like structures. At moderate particle densities and strong enough 
activities, active vesicles show dynamic highly-branched tethers filled with string-like arrangements of ABPs, which do not occur in the absence 
of particle adhesion to the membrane. At large volume fractions of ABPs, vesicles fluctuate for moderate particle activities, and elongate and finally split into two vesicles
for large ABP propulsion strengths. We also analyze membrane tension, active fluctuations, and ABP characteristics (e.g., mobility, clustering), 
and compare them to the case of active vesicles with non-adhesive ABPs. The adhesion of ABPs to the membrane significantly alters the behavior 
of active vesicles, and provides an additional parameter for controlling their behavior.  
\end{abstract}
\maketitle

\section{Introduction}
In recent years, there has been a growing interest in a variety of active matter systems which operate far from equilibrium and show rich dynamical behaviors and 
functions \cite{Ramaswamy_MAM_2010,Juelicher_HTA_2018,Gompper_MAM_2020,Shankar_TAM_2022}. Examples include biological systems ranging from cells to tissues \cite{Kelkar_MAC_2020,Trepat_MPP_2018}, collections of micro-swimmers \cite{Elgeti_PMS_2015,Bechinger_APC_2016}, and active engineered systems 
\cite{Needleman_AMI_2017,Banerjee_ACS_2020}. The growing research interest has been nurtured by rapid developments in microscale and nanoscale technologies which 
already allow for a well-controlled construction of complex multicomponent active systems and materials \cite{Needleman_AMI_2017,Schwille_MaxSynBio_2018,Bernheim_LMM_2018}. 
A prominent example is cell-mimicking systems, which are generally constructed from cell-based biological constituents, and include active nematics made of driven 
biofilaments \cite{Keber_TDV_2014,Duclos_TSD_2020}, and growing and dividing droplet-based or vesicle-based compartments \cite{Weirich_SOM_2019,Steinkuehler_CDV_2020}. 
In many other examples, biological materials are combined with active synthetic constituents with a hope to mimic various biological systems or even go beyond 
their functionality \cite{Elani_ILS_2021,vutukuri2020active}. Here, an interesting example is a closed membrane enclosing biological micro-swimmers such as bacteria \cite{takatori2020active,Le_Nagard_EBV_2022,park2022response} or synthetic self-propelled particles \cite{vutukuri2020active,Paoluzzi_SDF_2016,iyer2022non,li2019shape,peterson2021vesicle}. 
Active components inside the soft confinement exert forces on the surface, leading to highly dynamic non-equilibrium shape changes which resemble certain processes 
in living cells such as the formation of filopodia and lamellipodia \cite{Kelkar_MAC_2020,mattila2008filopodia,krause2014steering}, and active shape fluctuations of 
the membrane \cite{Tuvia_CMF_1997,Park_PNAS_10,Turlier_EPB_2016}.  

The main features that differentiate active vesicles from various membrane structures in equilibrium \cite{Seifert_STV_1991,Lipowsky_STM_2013} are active force 
generation due to the enclosed active components and dynamic shape changes of the membrane. For instance, swimming bacteria or motile synthetic particles within 
a vesicle induce the formation of tethers and protrusions which dynamically elongate and retract \cite{vutukuri2020active,takatori2020active,Le_Nagard_EBV_2022}. In equilibrium, 
string-of-pearls-like and tubular protrusions can be formed by amphipathic peptides or BAR domain proteins \cite{Lipowsky_STM_2013,Lipowsky_MSV_2022}, but these 
structures are static and correspond to a minimum of total energy. Therefore, different physical mechanisms govern the formation of various membrane structures 
in equilibrium and in non-equilibrium active vesicles. In particular, the curvature-induced clustering of active particles 
\cite{Fily_DSP_2014,fily2015dynamics,Iyer_MIPS_2022} at the membrane leads to the concentration of active forces at spots with a high curvature. 
Moreover, there exists a positive feedback mechanism between the induction of strong curvature by active particles and their clustering in places with large 
curvature, so that the shape of active vesicles is altered dynamically and collectively \cite{vutukuri2020active,iyer2022non}. Furthermore, active components 
within a vesicle give rise to a significant active tension due to the swim pressure exerted by the particles \cite{Takatori_SPG_2014}.

Apart from active forces, the deformation of a membrane can also occur as a consequence of adhesive interactions between the membrane and enclosed particles 
\cite{lipowsky1998vesicles,deserno2003wrapping,dasgupta2014shape,raatz2014cooperative,Dasgupta_NMP_2017}. In particular, adhesive interactions result in partial or full wrapping 
of the particles by the membrane, which can significantly reduce the force required for tether formation. Furthermore, the adhesion of multiple particles to 
the membrane often induces membrane-mediated interactions between the particles, leading to a cooperative wrapping of particles by the membrane 
\cite{raatz2014cooperative} and the formation of various particle structures at the membrane surface
\cite{koltover1999membrane,vsaric2012mechanism,vsaric2012fluid,cardellini2022membrane,bahrami2018curvature}. These interactions can enhance or reduce the clustering 
of active particles, potentially altering the behavior of active vesicles. In addition, it is plausible to expect that the adhesive interactions between 
the particles and the membrane can facilitate the existence of active forces away from the membrane (i.e., pulling forces), which is not possible for 
non-adhesive active particles which exert pushing forces toward the membrane. Finally, adhesive interactions of particles and pathogens with a 
cell membrane are essential for a variety of biological processes such as membrane translocation, viral budding, and phagocytosis
\cite{canton2012endocytosis,tzlil2004statistical,aderem1999mechanisms,rossman2011influenza}. 

In our study, we investigate numerically the combined effect of particle activity and adhesive interactions on the behavior of active vesicles. Fluid membrane vesicles 
are modeled as dynamically triangulated surfaces \cite{gompper2004triangulated,Kroll_CFM_1992} enclosing a number of active Brownian particles (ABPs). Adhesive 
interactions between the ABPs and the membrane are incorporated through the Lenard-Jones potential, whose strength is varied to induce various degrees of 
ABP wrapping by the membrane. A phase diagram of dynamic vesicle shapes is constructed as a function of the ABP propulsion strength and the volume fraction 
of particles within the vesicle. The presence of ABP adhesion to the membrane leads to qualitative changes in the phase diagram in comparison to that for 
non-adhesive ABPs \cite{vutukuri2020active}. For a weak particle activity, the adhesion interactions dominate, yielding nearly static vesicle shapes, which 
are similar to those in equilibrium with only adhesive interactions present. For moderate particle activities and volume fractions, complex tether structures 
filled with string-like arrangements of ABPs are formed, and characterized by a number of branching points. In contrast, for non-adhesive ABPs, the formed tethers 
show no significant branching, and the ABPs generally cluster at the end of membrane tethers \cite{vutukuri2020active,iyer2022non}. Finally, for a strong 
particle propulsion, active forces from the ABPs dominate over the adhesion interaction, and the resulting behavior of active vesicles is similar to those with 
non-adhesive ABPs. Also, membrane properties of the active vesicles and the characteristics of ABP clustering and mobility are 
analysed and compared to those of non-adhesive ABPs.    

The article is organized as follows. Section \ref{sec:model} provides all necessary details about the employed methods and models, including the parameters used 
in simulations. Section \ref{sec:diag} presents dynamic shape diagrams for two strengths of the ABP adhesion to the membrane. Membrane tension and the importance 
of ABP adhesion are discussed in Section \ref{sec:tension_res}. Vesicle shape fluctuations are analysed in Section \ref{sec:fluct}, and ABP characteristics are 
presented in Section \ref{sec:abp_prop}. Finally, we conclude in Section \ref{sec:concl}.

	\begin{table*}[th!]
		\centering
		\setlength{\tabcolsep}{12pt}
		\begin{tabular}{lll}
			\hline Parameters & Model units & Physical units \\
			\hline \multicolumn{3}{c} { \textbf{Principal properties }}  \\
			\hline vesicle radius in equilibrium $R$ & 32 & $8 \mu \mathrm{m}$ \\
			thermal energy unit $k_{\mathrm{B}} T$ & $0.2$ & $4.14 \times 10^{-21} \mathrm{~J}$ \\
			time scale $\tau=\gamma_{\mathrm{p}} R^{2} / \kappa_{\mathrm{c}}$ & $1.28 \times 10^{5}$ & $7.3 \mathrm{~s}$ \\
			\hline \multicolumn{3}{c} {\textbf{ Investigated properties }} \\
			\hline Peclet number $\mathrm{\text{Pe}}=\sigma \mathrm{v}_{\mathrm{p}} / \mathrm{D}_{\mathrm{t}}$ & $0-400$ & $0-400$ \\
			total number of SPPs $N_{\mathrm{p}}$ & $30-1458$ & $30-1458$ \\
			SPP volume fraction $\phi=N_{\mathrm{p}}(\sigma / 2 R)^{3}$ & $9 \times 10^{-3}-3.56 \times 10^{-1}$ & $9 \times 10^{-3}-3.56 \times 10^{-1}$ \\
			SPP diameter $\sigma$ & $R / 8$ & $1 \mu \mathrm{m}$  \\
			
			\hline \multicolumn{3}{c} { \textbf{Vesicle properties }} \\
			\hline number of vertices $N_{\mathrm{v}}$ & 30,000 & 30,000 \\
			bending rigidity $\kappa_{\mathrm{c}}$ & $20 k_{\mathrm{B}} T$ & $8.28 \times 10^{-20} \mathrm{~J}$ \\
			average bond length $l_{\mathrm{b}}$ & $4 R \sqrt{\frac{\pi}{N_{\mathrm{t}} \sqrt{3}}}$ & $0.176 \mu \mathrm{m}$ \\
			bond stiffness $k_{\mathrm{b}}$ & $80 k_{\mathrm{B}} T$ & $3.31 \times 10^{-19} \mathrm{~J}$ \\
			minimum bond length $l_{\min }$ & $0.6 l_{\mathrm{b}}$ & $0.11 \mu \mathrm{m}$ \\
			potential cutoff length $l_{\text{c}_1}$ & $0.8 l_{\mathrm{b}}$ & $0.14 \mu \mathrm{m}$ \\
			potential cutoff length $l_{\mathrm{c}_0}$ & $1.2 l_{\mathrm{b}}$ & $0.21 \mu \mathrm{m}$ \\
			maximum bond length $l_{\max }$ & $1.4 l_{\mathrm{b}}$ & $0.25 \mu \mathrm{m}$ \\
			desired vesicle area $A_0$ & $4 \pi R^{2}$ & $8.04 \times 10^{2} \mu \mathrm{m}^{2}$ \\
			local area stiffness $k_{\mathrm{a}}$ & $6.43 \times 10^{6} k_{\mathrm{B}} T / A$ & $3.3 \times 10^{-5} \mathrm{~J} / \mathrm{m}^{2}$ \\
			friction coefficient $\gamma_{\mathrm{m}}$ & $0.4 k_{\mathrm{B}} T \tau / R^{2}$ & $1.9 \times 10^{-10} \mathrm{~J} \cdot \mathrm{s} \cdot \mathrm{m}^{-2}$ \\
			flipping frequency $\omega$ & $6.4 \times 10^{6} \tau^{-1}$ & $8.8 \times 10^{5} \mathrm{~s}^{-1}$ \\
			flipping probability $\psi$ & $0.3$ & $0.3$\\
			\hline \multicolumn{3}{c} { \textbf{ABP properties } }\\
			\hline translational friction $\gamma_{\mathrm{p}}$ & $20 k_{\mathrm{B}} T \tau / R^{2}$ & $9.4 \times 10^{-9} \mathrm{~J} \cdot \mathrm{s} \cdot \mathrm{m}^{-2}$ \\
			translational diffusion $D_{\mathrm{t}}$ & $k_{\mathrm{B}} T / \gamma_{\mathrm{p}}$ & $4.4 \times 10^{-13} \mathrm{~m}^{2} \cdot \mathrm{s}^{-1}$ \\
			rotational diffusion $D_{\mathrm{r}}$ & $3 D_{\mathrm{t}} / \sigma^{2}$ & $1.32 \mathrm{~s}^{-1}$ \\
			LJ potential depth (adhesion energy) $\epsilon$ & $2.5 k_{\mathrm{B}} T-3.5 k_{\mathrm{B}} T$ & $1.03 \times 10^{-20}-1.45 \times 10^{-20} \mathrm{~J}$ \\
			 \\
		\end{tabular}
		\caption{Parameters used for simulations of vesicles enclosing adhesive ABPs both in model and physical units. $N_{\mathrm{t}} = 2N_{\mathrm{v}} - 4$ 
		is the number of triangular faces in the vesicle discretization.}
		\label{tab:param}
	\end{table*}

\section{Methods and models}
\label{sec:model}

An active vesicle is represented by a closed fluid membrane of spherical topology with radius $R$, enclosing $N_p$ active Brownian particles (ABPs). 
The activity of the particles is described by the dimensionless Peclet number $\text{Pe}=\sigma v_{\mathrm{p}}/D_{\mathrm{t}}$, where $\sigma$ is the particle 
diameter, $v_{\mathrm{p}}$ is the propulsion velocity, and $D_{\mathrm{t}}$ is the translational diffusion coefficient. Note that $\text{Pe}$ is a measure of 
the propulsion force $f_{\mathrm{p}}$ of ABPs, with $v_{\mathrm{p}}=f_{\mathrm{p}}/\gamma_{\mathrm{p}}$ and $D_{\mathrm{t}} = k_{\mathrm{B}}T/\gamma_{\mathrm{p}}$, 
where $\gamma_{\mathrm{p}}$ is the translational friction coefficient, so that $\text{Pe}=f_{\mathrm{p}} \sigma/k_{\mathrm{B}}T$. Particle volume fraction 
within the vesicle is given by $\phi = N_{\mathrm{p}} (\sigma/2R)^3$. Table \ref{tab:param} presents 
all simulation parameters. 
		
\subsection{Model of adhesive active Brownian particles}
	
ABPs are modeled as active spherical particles without hydrodynamic interactions. Each ABP experiences a propulsion force $f_{\mathrm{p}}$ that acts along 
an orientation vector $\boldsymbol{e}_i$. The force results in a propulsion velocity $ v_{\mathrm{p}} = f_{\mathrm{p}}/\gamma_{\mathrm{p}}$. 
The orientation vector $\boldsymbol{e}_i$ is subject to orientational diffusion $\dot{\boldsymbol{e}}_i =\boldsymbol{\zeta}_i \times \boldsymbol{e}_i$, where 
$\boldsymbol{\zeta}_i$ is a Gaussian random process with $\langle\boldsymbol{\zeta}_i(t)\rangle=0$ and $ \langle\boldsymbol{\zeta}_i(t)
\boldsymbol{\zeta}_j(t')\rangle = 2D_{\mathrm{r}}\delta_{ij}\delta(t-t')$ with a rotational diffusional coefficient $D_{\mathrm{r}}$. $D_{\mathrm{r}}$ 
is related to the ABP size $\sigma$ and translational diffusion coefficient $D_{\mathrm{t}}$ as $D_{\mathrm{t}}= D_{\mathrm{r}}\sigma^2/3$. The ABPs repel 
each other, which is implemented through the repulsive part of the 12-6 Lennard-Jones (LJ) potential, with the potential minimum and cut-off at 
$r_{\mathrm{m}}^{\mathrm{p-p}}=r_{\mathrm{c}}^{\mathrm{p-p}} = 2^{1/6}\sigma$ for ABP-ABP interactions. Furthermore, the ABPs are attracted to the membrane, 
which is implemented by the full 12-6 LJ potential with a minimum at $r_{\mathrm{m}}^{\mathrm{p-m}} = 2^{1/6}\sigma/2$. The potential cut-off for 
ABP-membrane interactions is set to $r_{\mathrm{c}}^{\mathrm{p-m}} = 2^{1/6} \sigma$. 
		
\begin{figure*}
    \centering
    \includegraphics{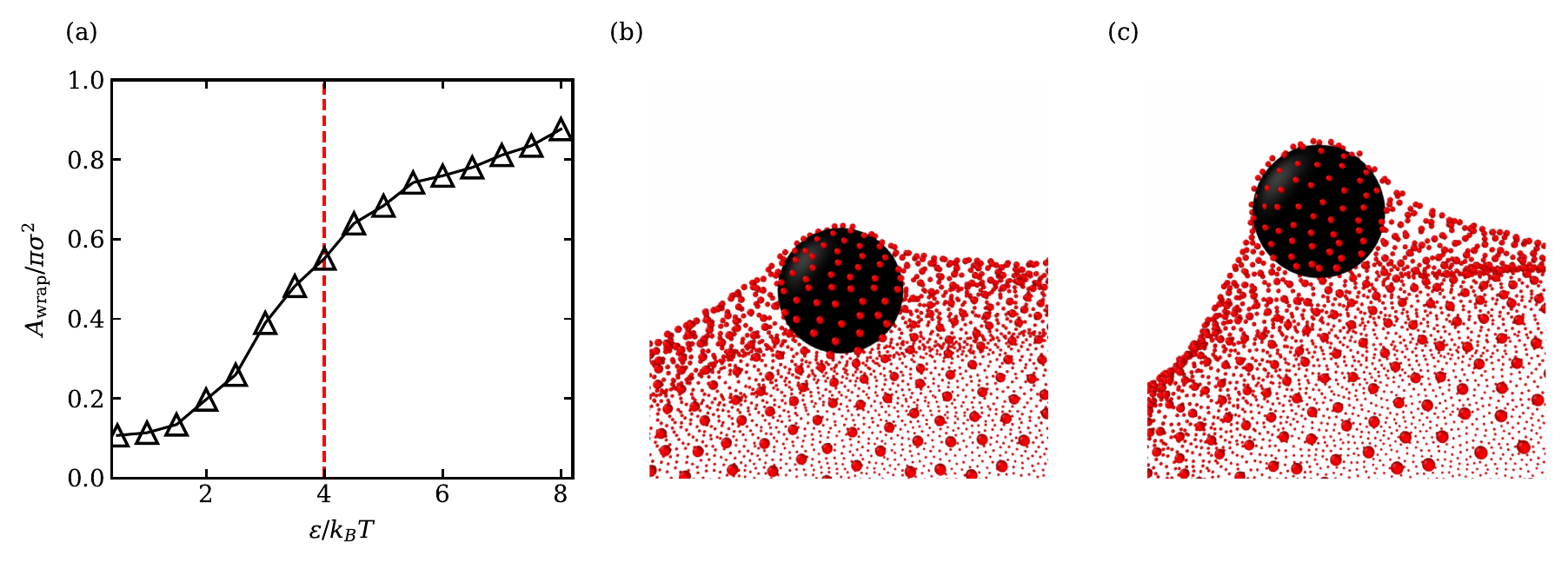}
    \caption{(a) Fraction of wrapped area $A_\text{wrap}$ of the ABP as a function of $\epsilon$. The dashed red line marks theoretical predictions of the critical 
    $\epsilon_c$ for the transition from unwrapped to fully wrapped state \cite{lipowsky1998vesicles}. Due to the long range of interactions between 
    the particle and the membrane, the transition is gradual and the particle is only partially wrapped at $\epsilon_c$ \cite{raatz2014cooperative}. 
    Partially wrapped states of a particle (black) by the membrane (red) for (b) $\epsilon=2.5k_BT$ and (c) $\epsilon=3.5k_BT$. }
    \label{fig:adhesion_strength}
\end{figure*}
		
\subsection{Membrane model}
	
The vesicle is modeled by a dynamically triangulated membrane of spherical topology consisting of  $N_{\mathrm{v}}$ linked vertices \cite{gompper2004triangulated,Kroll_CFM_1992}. 
The interaction between linked vertices is controlled via a tethering potential \cite{gompper2004triangulated,Noguchi_DFV_2005} that is a combination 
of attractive and repulsive parts
\begin{equation}
	U_{\text{att}}(r)=\begin{cases}
		k_{\mathrm{b}} \dfrac{\exp[1/(l_{{\mathrm{c_0}}}-r)]}{l_{{\mathrm{max}}}-r}, & \text{if $r>l_{{\mathrm{c_0}}},$}\\
		0, & \text{if $r\leq l_{{\mathrm{c_0}}},$}
	\end{cases}
\end{equation}
	
\begin{equation}
	U_{\text{rep}}(r)=\begin{cases}
		k_{\mathrm{b}} \dfrac{\exp[1/(r-l_{{\mathrm{c_1}}})]}{r-l_{{\mathrm{min}}}}, & \text{if $r<l_{{\mathrm{c_1}}},$}\\
		0, & \text{if $r\geq l_{{\mathrm{c_1}}}.$}
	\end{cases}
\end{equation}
Here, $k_{\mathrm{b}}$ is the bond stiffness, $l_{{\mathrm{min}}}$ and $l_{{\mathrm{max}}}$ are the minimum and maximum bond lengths, and 
$l_{{\mathrm{c_0}}}$ and $l_{{\mathrm{c_1}}}$ are the potential cutoff lengths.
	
The membrane bending elasticity is modeled by the Helfrich curvature energy \cite{Helfrich_EPB_1973},
\begin{equation}
	U_{{\mathrm{bend}}} = 2\kappa_{\mathrm{c}} \oint \bar{c}^2 dA,
	\label{helfirch_eq}
\end{equation}
where $\kappa_{\mathrm{c}}$ is the bending rigidity and $\bar{c}=(c_1 + c_2)/2$ is the mean local curvature at the membrane surface element $dA$.  In the discretized 
form, it becomes \cite{Gompper_1996,gompper1997networks}
\begin{equation}
	U_{\text {bend}}=2\kappa_{\mathrm{c}} \sum_{i=1}^{N_{\mathrm{v}}}\sigma_{i}\bar{c}^2_i,
	\label{helfirch_eq_des}
\end{equation}
where $\bar{c}_i = \mathbf{n}_{i} \cdot \sum_{j(i)} \sigma_{i j} \mathbf{r}_{i j}/(2 \sigma_{i}r_{i j})$ is the discretized mean curvature at 
vertex $i$,  $\mathbf{n}_{i}$ is the unit normal at the membrane vertex $i$, $\sigma_{i} =\sum_{j(i)} \sigma_{i j}r_{i j}$ is the area corresponding 
to vertex $i$ (the area of the dual cell), $j(i)$ corresponds to all vertices linked to vertex $i$, and $\sigma_{i j}=r_{ij}(\cot\theta_1+ \cot\theta_2)/2$ 
is the length of the bond in the dual lattice, where $\theta_1$ and $\theta_2$ are the angles at the two vertices opposite to the edge $ij$ in the dihedral. 
In practice, since the dihedral terms corresponding to $\sigma_{ij}$ are additive, the local curvature at each vertex can be calculated by summing 
over contributions from all triangles containing that vertex. 
	
The area conservation is imposed locally to each triangle by the potential 
\begin{equation}
	U _{\text A}= \frac{k_{\mathrm{a}}}{2}\sum_{i=1}^{N_{\mathrm{t}}} \frac{(A_i-A_{\mathrm{l}})^2}{ A_{\mathrm{l}}},
\end{equation}
where $N_{\mathrm{t}}=2(N_{\mathrm{v}} -2)$ is the number of triangles, $A_{\mathrm{l}} = A_0/N_{\mathrm{t}}$ is the targeted local area ($A_0$ is 
the total membrane area), $A_i$ is the instantaneous local area,  and $k_{\mathrm{a}}$ is the local-area conservation coefficient. We do not impose 
any volume constraints, and therefore, the vesicle volume is free to change. 
	
Membrane fluidity is modelled by a stochastic flipping of bonds following a Monte-Carlo scheme. The bond shared by each pair of adjacent triangles 
can be flipped to connect the two previously unconnected vertices \cite{gompper2004triangulated,gompper1997networks}. The flipping is performed with 
a frequency $\omega$ and probability $\psi$. An energetically favorable bond flip is accepted with a probability of $p=1$. For an energetically unfavorable 
flip, the resulting change in energy due to an attempted bond flip $\Delta U = \Delta U_{{\mathrm{att}}} + \Delta U_{{\mathrm{rep}}} + \Delta U_{\mathrm{A}}$ 
determines the probability of the flipping as $p= \exp[-\Delta U/k_{\mathrm{B}} T]$. The resulting membrane fluidity can be characterized by a 2D membrane 
viscosity for the selected frequency $\omega$ and flipping probability $\psi$ \cite{Noguchi_DFV_2005,noguchi2004fluidves}.
	
\subsection{Equation of motion}
	
The system evolves in time according to the Langevin equation 
\begin{equation}
	m \ddot{\mathbf{r}}_i =-\nabla_i U_{\text {tot}}-\gamma \dot{\mathbf{r}}_i + \sqrt{2 \gamma k_{\mathrm{B}} T} \xi_i(t),
\end{equation}
where $m$ is the mass of membrane particle or ABP, $\ddot{\mathbf{r}}_{i}$ and $\dot{\mathbf{r}}_{i}$ represent the second and first time derivatives 
of particle positions, $\nabla_i$ is the spatial derivative at particle $i$, and $U_{{\mathrm{tot}}}$ is the sum of all interaction potentials described 
above. The effect of a viscous fluid is mimicked by the friction co-efficient $\gamma$, whose value can be different for membrane particles and ABPs, 
see Table~\ref{tab:param}. Thermal fluctuations are modelled as a Gaussian random process $\xi_{i}$ with $\langle\xi_{i}(t)\rangle=0$ and 
$\langle\xi_{i}(t)\xi_{j}(t')\rangle = \delta_{ij}\delta(t-t')$. Inertial effects are minimized by performing the simulations in the over-damped limit 
with $m$ and $\gamma$ such that $\tau_\text{t}=m/\gamma<1\ll \tau_\text{r} = D^{-1}_\text{r}$. The positions and velocities of all particles are 
integrated using the velocity-Verlet algorithm \cite{Allen_CSL_1991}.
	
\begin{figure*}[t!]
    \centering
    \includegraphics[scale=0.9]{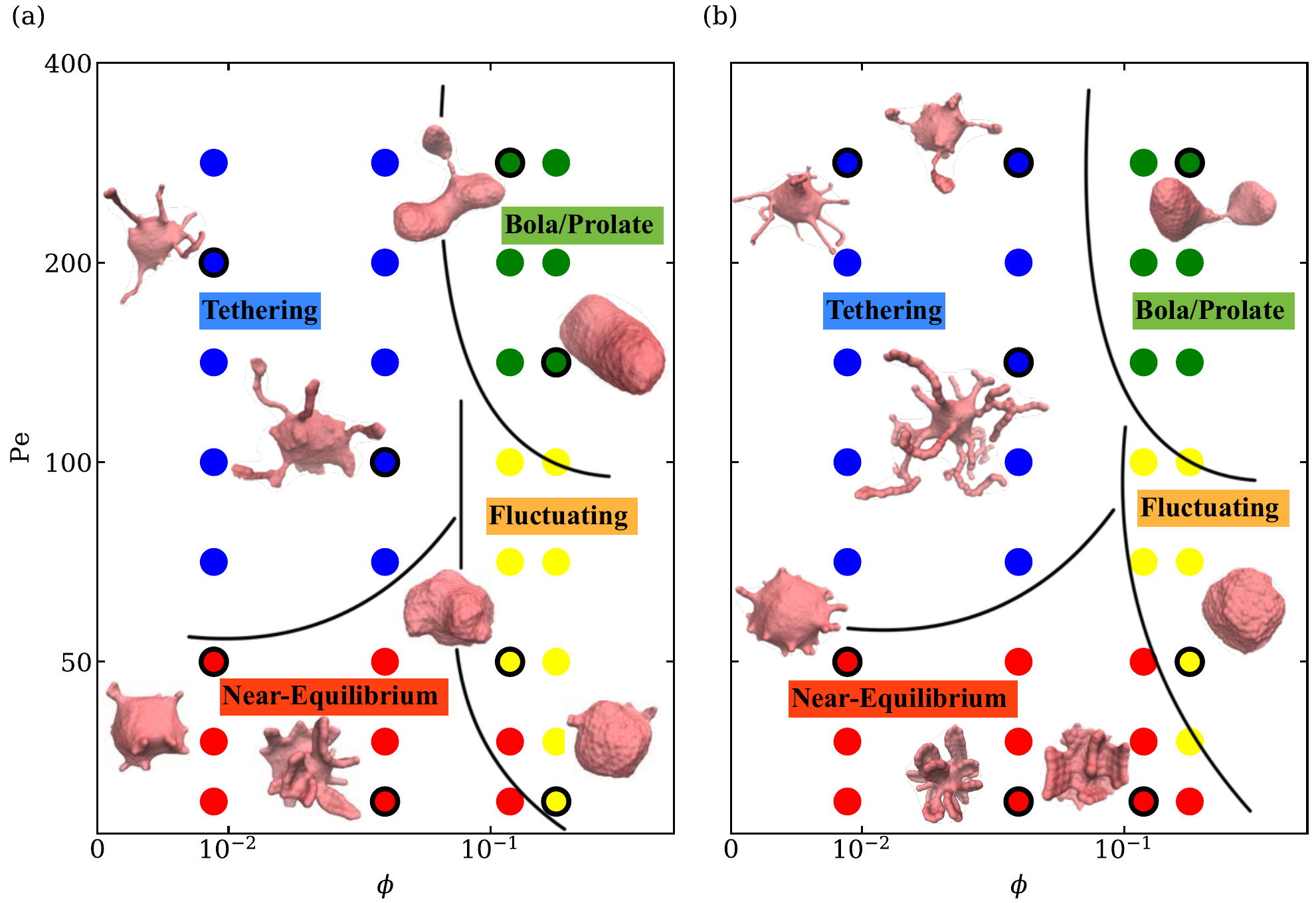}
    \caption{Phase diagrams of vesicle-shape changes as a function of $\text{Pe}$ and $\phi$ for two different adhesion strengths 
    (a) $\epsilon = 2.5 k_BT$ and (b) $\epsilon = 3.5 k_BT$. Four regions are observed, including the tethering (blue symbols), 
    bola/prolate (green symbols), fluctuating (yellow symbols), and near-equilibrium (red symbols) regimes. The points corresponding to 
    the displayed snapshots have black outlines. The black lines provide an approximate demarcation of the different regimes, serving as a guide 
    to the eye. For a visual illustration of dynamic shape changes of active vesicles, see also Movies S1-S4.}
    \label{fig:phase_eps}
\end{figure*}
	
\subsection{ABP adhesion and membrane wrapping}

Adhesion interactions between the ABPs and the membrane are mediated by the LJ potential whose strength is characterized by the potential depth 
$\epsilon$. Adhesion strength determines the degree of particle wrapping by the membrane, with energetic costs due to membrane bending and tension. 
The ratio of the membrane bending modulus $\kappa$ and the lateral tension $\lambda$ defines a length $\tilde{l}= \sqrt{\kappa/\lambda}$, below which membrane deformations are mainly controlled by the bending energy, while deformations on length scales larger 
than $\tilde{l}$ are dominated by tension \cite{deserno2003wrapping}. If tension is neglected and the membrane covers area $A_\text{wrap} \leq \pi \sigma^2$ of the particle 
(e.g.,~partial wrapping), the adhesion ($E_{\text{ad}}$) and bending ($E_{\text{bend}}$) energies are given by 
\begin{equation}
	E_{\text{ad}} = -\omega A_\text{wrap}, \quad \quad E_{\text{bend}}= 8 \kappa A_\text{wrap} / \sigma^2,
\end{equation}
where $\omega$ is the adhesion strength per unit area. In this case, the minimum of the total energy corresponds to complete wrapping of the particle 
by the membrane (i.e., $A_\text{wrap} = \pi \sigma^2$), which occurs for $\omega > \omega_\text{min} = 8\kappa/\sigma^2$ \cite{lipowsky1998vesicles}. 
Therefore, in the absence of membrane tension, the particle is in an unwrapped state for $\omega < \omega_\text{min}$, while the particle is fully 
wrapped for $\omega > \omega_\text{min}$ with no energy barrier to overcome. However, in the presence of tension, particle adhesion shows a continuous 
transition from the unwrapped to partially wrapped state at $\omega_\text{min}$, while the transition to the fully wrapped state is discontinuous and 
has an energy barrier \cite{deserno2003wrapping}. 

To relate the adhesion strength $\omega$ per unit area and the strength $\epsilon$ of the LJ potential in simulations, we consider the attraction 
of a single membrane vertex to an ABP, such that $\epsilon = 2 \omega A_\mathrm{l}$ with $2A_\mathrm{l}$ being the area of the vertex. For the parameters 
in Table~\ref{tab:param}, the transition from the unwrapped to a wrapped state is expected at $\omega_\text{min} = 8\kappa/\sigma^2$ which implies 
$\epsilon_c \simeq 4k_BT$. In our simulations, adhesive interactions between ABPs and the membrane are exerted up to a distance of $\sigma/2$ from 
the ABP surface, and are therefore long ranged. Theoretical predictions of particle wrapping for long-ranged adhesive interactions indicate that 
the transition to the fully wrapped state is gradual \cite{raatz2014cooperative}, which is consistent with the area $A_\text{wrap}$ of particle 
wrapping as a function of $\epsilon$ shown in Fig.~\ref{fig:adhesion_strength}(a). Thus, the fully wrapped state requires adhesion interactions 
with $\epsilon > \epsilon_c$. For further simulations, we have selected two adhesion strengths of $\epsilon=2.5k_BT$ and $\epsilon=3.5k_BT$, which 
correspond to a moderate degree of wrapping illustrated in Fig.~\ref{fig:adhesion_strength}(b,c). 

\section{Results}
\label{sec:results}

\subsection{Dynamic phase diagram}
\label{sec:diag}

\begin{figure*}[ht!]
    \centering
    \includegraphics{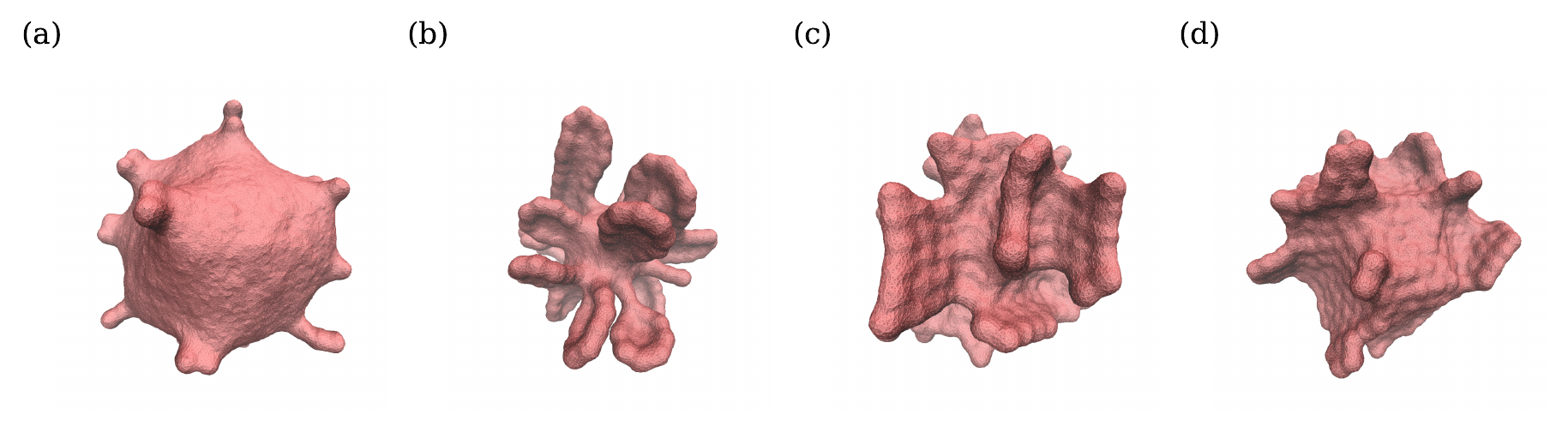}
    \caption{Vesicle shapes in the near-equilibrium regime at $\text{Pe}=15$ and $\epsilon=3.5k_BT$ for (a) $\phi=0.009$, (b) $\phi=0.04$ (see Movie S4), 
    (c) $\phi=0.12$, and (d) $\phi=0.18$. Different nearly-frozen structures of the ABPs are observed, including ring-like and sheet-like 
    arrangements.}
    \label{fig:eq_sim}
\end{figure*}

Figure \ref{fig:phase_eps} presents phase diagrams of dynamic shape changes of active vesicles as a function of $\text{Pe}$ and $\phi$ for two different 
adhesion strengths $\epsilon$ (see also Movies S1-S4). At small $\text{Pe} \lesssim 50$, the formation of buds for low particle densities, ring-like aggregates of ABPs 
for intermediate $\phi$ values, and ABP aggregates with a hexagonal closed-packed (HCP) structure for large $\phi$ are observed and illustrated 
in Fig.~\ref{fig:eq_sim} for $\text{Pe}=15$. Some of these structures have previously been observed in studies of passive particles adhering to 
a membrane \cite{koltover1999membrane,vsaric2012mechanism,vsaric2012fluid,cardellini2022membrane}. Furthermore, for $\text{Pe} \lesssim 50$, ABPs adhered 
to the membrane show little dynamics, suggesting that adhesive forces dominate over particle activity. As a result, active vesicles for 
$\text{Pe} \lesssim 50$ are close to an equilibrium state, with ABP aggregate structures similar to those of equilibrium systems at $\text{Pe}=0$.  

In the near-equilibrium regime at low ABP densities, both individually wrapped particles and short strings of several ABPs within membrane tubes 
[see Fig.~\ref{fig:eq_sim}(a)] are observed due to the competition between repulsive curvature-mediated interactions \cite{bahrami2018curvature} 
and the in-plane motion from ABP propulsion. As the ABP volume fraction is increased, strong cooperative wrapping of ABPs is observed, as shown 
in Fig.~\ref{fig:eq_sim}(b-c). Here, it is likely that the gain in energy due to the cooperative wrapping overcomes the curvature-mediated repulsion. 
Cooperative wrapping of several particles is also enhanced by the interaction range of an adhesion potential \cite{raatz2014cooperative}. 
Furthermore, the vesicle is free to change its volume in our simulations, and therefore, the area fraction of adhered membrane can be large.
As a result, extreme deformations of the vesicle with protruding ring-like and sheet-like structures are observed and illustrated 
in Fig.~\ref{fig:eq_sim}(b-c). For the largest volume fraction of ABPs ($\phi = 0.18$), membrane deformations are reduced [see Fig.~\ref{fig:eq_sim}(d)] 
in comparison to the cases of $\phi = 0.04$ and $\phi=0.12$, because the gain in adhered membrane area is restricted at some point by the volume of 
the ABP content. Therefore, if the vesicle volume were constrained to near-spherical values, membrane deformations are expected to be reduced, since the gain in adhered membrane 
would be restricted by an increase in membrane tension due to the constrained vesicle volume. Note that for the lower adhesion strength of 
$\epsilon=2.5k_BT$, membrane deformations are less pronounced than in the case of $\epsilon=3.5k_BT$ (see Fig.~\ref{fig:phase_eps} for 
$\text{Pe} \lesssim 50$) due to the competition between adhesion and bending energies.    

\begin{figure*}[ht!]
    \centering
    \includegraphics[scale=0.9]{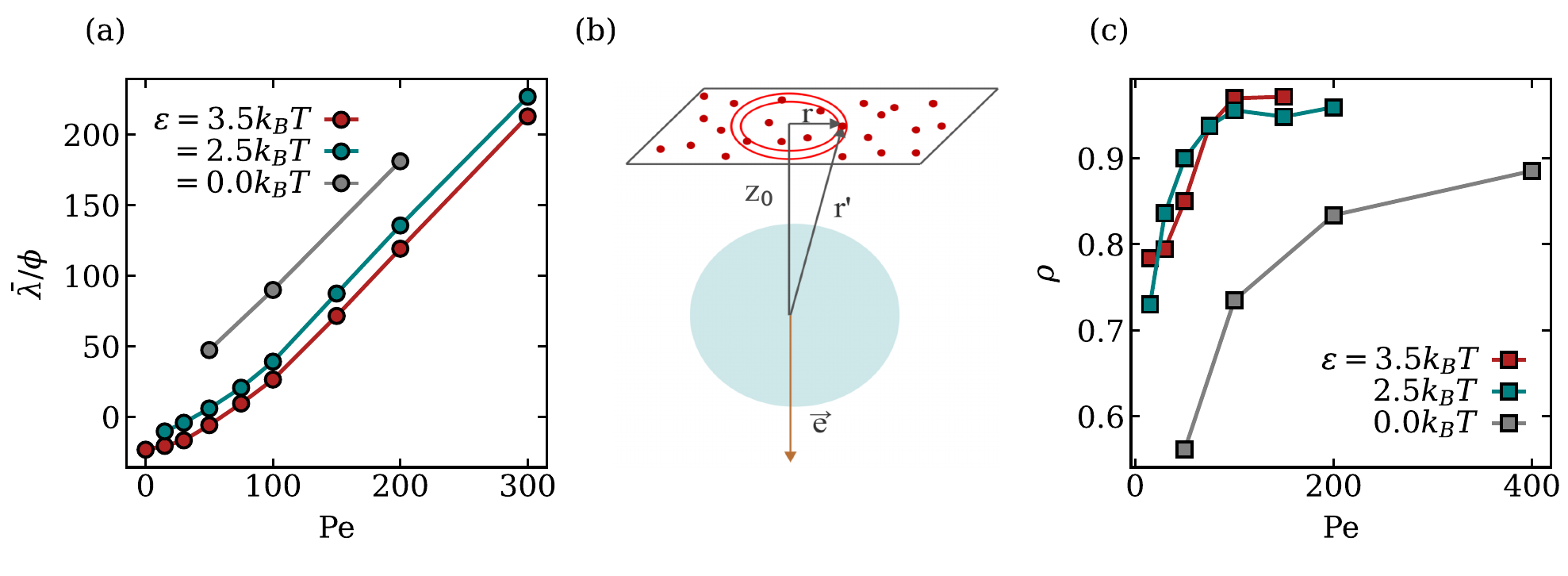}
    \caption{ (a) Mean local vesicle tension $\bar{\lambda}$ as a function of $\text{Pe}$ for different adhesion strengths $\epsilon$ at $\phi=0.18$. 
    (b) Sketch of an ABP (blue) interacting with a flat membrane at a distance $z_0$. Membrane vertices are depicted in red and 
    the orientation vector $\vec{e}$ of the ABP is pointing away from the membrane. (c) Mean fraction $\rho$ of the ABPs in contact with the membrane 
    as a function of $\text{Pe}$ for different $\epsilon$ values at $\phi=0.18$. }
    \label{fig:tension_rho}
\end{figure*}
 
As $\text{Pe}$ is increased, ABP propulsion starts to dominate over the adhesive forces, and the non-equilibrium nature of active vesicles becomes apparent. 
At low particle densities ($\phi \lesssim 0.07$), ABP activity leads to the formation of dynamic tether-like structures, which are filled by 
string-like arrangements of ABPs. This behavior is qualitatively different from the tether formation by ABPs in the absence of adhesive interactions, 
where particle clustering takes place at the end of a tether \cite{vutukuri2020active,iyer2022non}. Note that the string-like arrangement of particles 
in membrane tubes is favored by long-ranged adhesive interactions \cite{raatz2014cooperative,agudo2016stabilization}. Another qualitative difference 
of the formed tethered structures by adhesive ABPs in comparison to those by non-adhesive active particles \cite{vutukuri2020active,iyer2022non} 
is that the tethered structures in Fig.~\ref{fig:phase_eps} are often highly branched. Since ABPs spend a considerable time in string-like configurations 
within membrane tethers, ABPs can change their orientation due to rotational diffusion and initiate branch formation from the existing tether. 
In the absence of adhesive interactions, ABPs quickly travel between the base of a tether and its end (or vise versa), and thus cannot easily initiate branched 
tethers \cite{vutukuri2020active,iyer2022non}. Therefore, adhesive interactions promote the formation of branched tether structures and stabilize them. 
At the lower adhesion strength of $\epsilon=2.5k_BT$, ABPs cluster more at tether ends than for the case of $\epsilon=3.5k_BT$, and result in less 
branched structures, as shown in Fig.~\ref{fig:phase_eps}. A similar effect is observed with increasing particle activity (or $\text{Pe}$), suggesting 
that branched tether structures and string-like arrangements of ABPs are indeed a consequence of particle adhesion to the membrane, which is lost 
when ABPs have a sufficient force to detach from the membrane. Note that the tethering regime for adhesive ABPs occurs at significantly lower $\text{Pe}$ 
numbers when compared to the non-adhesive ABP case \cite{vutukuri2020active,iyer2022non} because particle adhesion facilitates wrapping, reducing 
the energy barrier required for the formation of tethers. 

At large particle densities ($\phi \gtrsim 0.07$) and for $\text{Pe}$ values beyond the near-equilibrium regime, a fluctuating phase first develops, 
where shape changes of the vesicle are moderate and resemble membrane fluctuations. In Section \ref{sec:fluct}, we will show that vesicle shape 
fluctuations for adhesive ABPs are different from those for the non-adhesive ABP case \cite{vutukuri2020active}. As $\text{Pe}$ is further increased 
for $\phi \gtrsim 0.07$, the ABPs form large clusters which can push in opposing directions and result in vesicle elongation or even splitting into two vesicles, similar 
to the non-adhesive ABP case \cite{vutukuri2020active}. Thus, the effect of adhesive interactions is prevalent only for low to intermediate $\text{Pe}$ 
values, where the adhesive forces are larger than or comparable to ABP propulsion forces. 

\subsection{Membrane tension}
\label{sec:tension_res}

The mean vesicle tension $\bar{\lambda}$ computed from local membrane stresses (see Appendix A) for different 
adhesion energies is shown in Fig.~\ref{fig:tension_rho}(a). For small $\text{Pe} \lesssim 50$ in the case of adhesive ABPs, the mean local tension of the vesicle is slightly negative, 
which indicates local compression of the membrane vertices. The local contraction of the membrane is facilitated by adhesive interactions, which are 
relatively long ranged, and favor the adhesion of more membrane vertices to the ABPs, leading to local compression of the membrane within the adhesion 
area. Furthermore, at small $\text{Pe}$, active particles generate a relatively low swim pressure at the membrane, so that the membrane tension 
remains slightly negative.    

Membrane tension $\bar{\lambda}$ for $\epsilon > 0$ in Fig.~\ref{fig:tension_rho}(a) exhibits two different regimes. For $\text{Pe} < 100$, 
the dependence of $\bar{\lambda}$ is non-linear, while for $\text{Pe} > 100$, $\bar{\lambda}$ increases linearly with increasing $\text{Pe}$, 
similar to the case of $\epsilon=0$. For active vesicles with non-adhesive ABPs, the linear growth in $\bar{\lambda}$ is determined by the swim 
pressure of ABPs on the membrane, such that $\bar{\lambda} / \lambda_0 = \chi \text{Pe}\phi$, where $\lambda_0=R^2k_BT/(\pi\sigma^4)$ is 
a normalization factor and $\chi$ is the active tension weight related to the alignment of propulsion direction with the membrane normal \cite{iyer2022non}. Therefore, the linear regime of $\bar{\lambda}$ for 
active vesicles with adhesive ABPs is also due to the swim pressure of ABPs on the membrane, because for large $\text{Pe}$, the ABP 
propulsion force dominates over adhesion interactions. However, the non-linear dependence of $\bar{\lambda}$ for $\text{Pe} < 100$ and 
$\epsilon > 0$ is due to the interplay of swim pressure and particle adhesion to the membrane.      

\begin{figure*}[htpb!]
    \centering
    \includegraphics{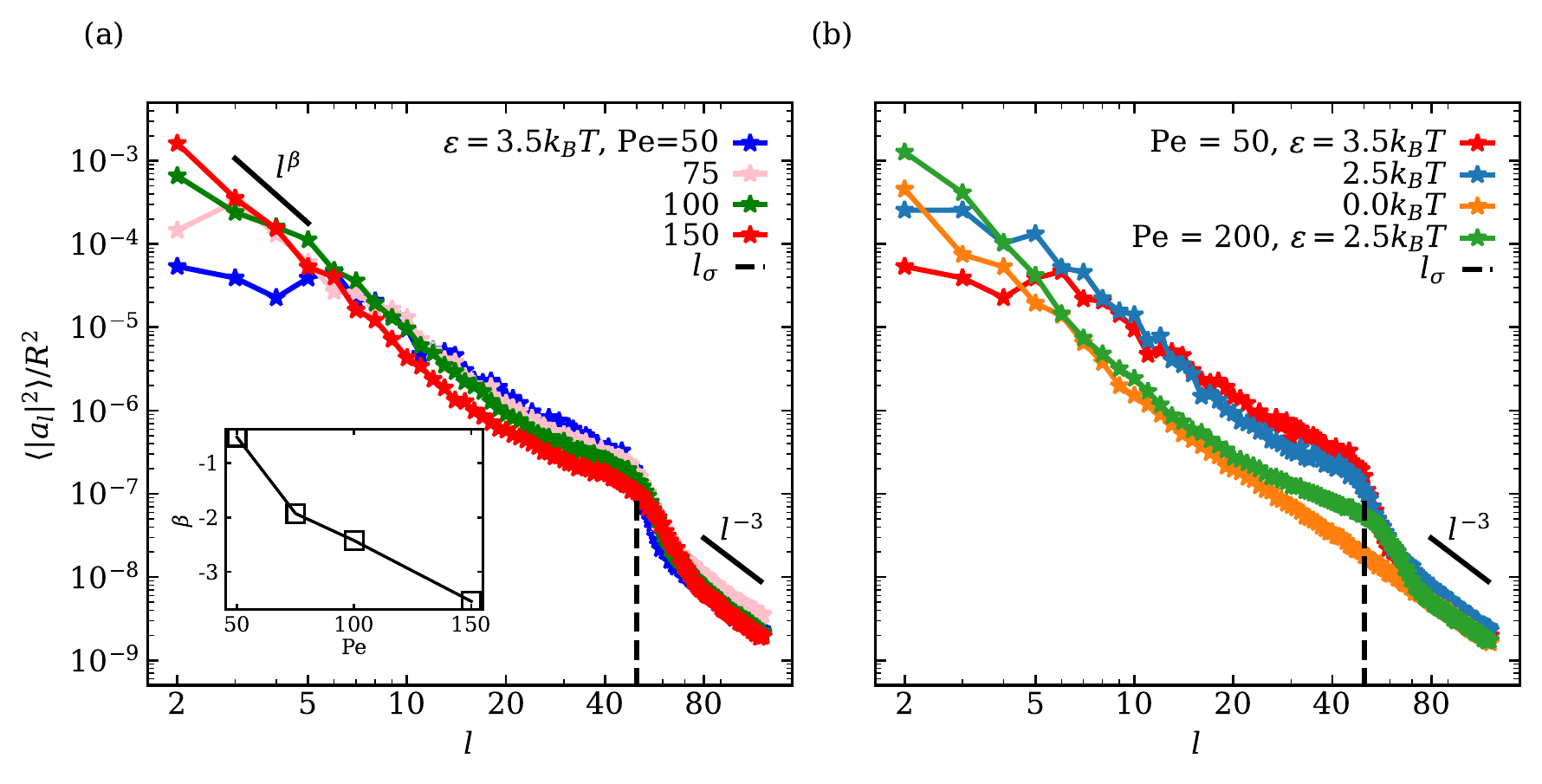}
    \caption{Mode spectra of vesicle-shape fluctuations at (a) $\epsilon=3.5 k_BT$ for different $\text{Pe}$ values and at (b) $\text{Pe}=50$ 
    for different $\epsilon$ values. Large wavelength (low mode) fluctuations are suppressed at low $\text{Pe}$ for a strong ABP adhesion, resulting in 
    a plateau-like region at $l \lesssim 10$. The inset in (a) shows the slope $\beta$ of low-mode fluctuations with increasing $\text{Pe}$. The dashed lines indicate the mode number $l_\sigma = 2\pi R / \sigma \simeq 50$, representing a wavelength of the ABP size.}
    \label{fig:fluct}
\end{figure*}

The location of the transition from the non-linear to the linear increase in $\bar{\lambda}$ with increasing $\text{Pe}$ can be estimated using a simple model, where 
an adhesive particle placed at a distance $z_0$ from a flat membrane attempts to escape the surface, see Fig.~\ref{fig:tension_rho}(b). The attractive force exerted on the particle due to the membrane patch at a distance $r'=\sqrt{r^2+z_0^2}$ with an area $2 \pi r dr$ is given by 
\begin{equation}
	dF = 24\epsilon \left [2 \left(\frac{\sigma}{2r'}\right)^{12} -  \left(\frac{\sigma}{2r'}\right)^{6} \right] n  \frac{z_0}{r'}\frac{2\pi r}{r'} dr,
\end{equation}
where $n=N_v/4\pi R^2$ is the number density of vertices at the membrane, and the factor $z_0/r'$ is due to the projection of the force onto 
the normal-to-the-surface direction. When the propulsion direction $\vec{e}$ of the particle points away from the membrane along the normal, 
force balance implies  
\begin{align}
    \begin{split}
    \text{Pe} & =\frac{\sigma}{k_BT}\int_{0}^{\sqrt{(r_{\mathrm{c}}^{\mathrm{p-m}})^2 - z_0^2}}dF \\
    & = \frac{8\pi\epsilon n\sigma z_0}{k_BT}	
    \left[\left( \frac{2r'}{\sigma} \right)^{-12}-\left( \frac{2r'}{\sigma} \right)^{-6}\right]_{z_0}^{r_{\mathrm{c}}^{\mathrm{p-m}}}.
    \end{split}
\end{align}
This expression allows the calculation of a maximum $\text{Pe}$ required for ABP detachment from the membrane, yielding
$\text{Pe}_{max}\approx 385$ for $\epsilon = 3.0 k_BT$ and $z_0\simeq1.14\sigma/2$ (i.e., $z_0$ is the distance from the flat membrane at which the maximum in $\text{Pe}$ is obtained). Here, the local 
curvature of the membrane is neglected, which would result in an increase of the detachment force. From simulations with a frozen membrane, the 
detachment of an ABP with $\epsilon = 3.0 k_BT$ takes place at $\text{Pe}_{max}\approx 390$, in good agreement with the analytical estimate. 
However, for a deformable membrane, thermal undulations lead to a steric repulsion of the ABP from the membrane \cite{helfrich1978steric}, which 
causes a decrease in the detachment force. The repulsive force exerted on the ABP by the fluctuating membrane can be estimated in terms of 
$\text{Pe}$ as \cite{bickel2003depletion,spanke2020wrapping}
\begin{equation}
    \text{Pe}_{noise} =\frac{\sigma}{k_BT} (2\pi R_{\text{eff}}) \frac{c(k_BT)^2}{\kappa h^2},
\end{equation}
where $R_{\text{eff}}=2^{1/6}\sigma/2$ is the effective radius of the particle (here, the equilibrium distance between the particle and membrane vertices), $h=z_0-R_{\text{eff}} \simeq 0.02 \sigma/2$ is the distance between the ABP surface and the membrane, and $c$ is a constant in the range $(0.01,0.23)$
\cite{helfrich1978steric,Dinsmore_HSV_1998,Gompper_SIS_1989,spanke2020wrapping}. Note that the magnitude of $h$ in our case is similar to the average fluctuation amplitude $\bar{h}_\sigma$ of a flat tensionless membrane on a length scale of the particle size $\sigma$, where $\bar{h}_\sigma = \sqrt{2k_B T/ \kappa /(2\pi)^3} \sigma/2 \simeq 0.02 \sigma/2$. The range of $c \in (0.01,0.23)$ corresponds to a broad range of $\text{Pe}_{noise} \in (20,500)$. 
Recent experiments of particle wrapping by a lipid membrane \cite{spanke2020wrapping} suggest a much narrower range of $c \in (0.03,0.06)$, 
corresponding to $\text{Pe}_{noise} \in (65,130)$. Taking the median value of $\text{Pe}_{noise} \simeq 100$ for $c=0.045$, $\text{Pe}$ required 
for the ABP detachment becomes $\text{Pe}_{detach} = \text{Pe}_{max} - \text{Pe}_{noise} \simeq 290$. From simulations of a single ABP adhered
to a fluctuating membrane with $\epsilon = 3.0 k_BT$, the detachment force corresponds to $\text{Pe} \simeq 200$. This $\text{Pe}$ value 
is in a reasonable agreement with the theoretical estimate of $\text{Pe}_{detach}$, taking into account that $\text{Pe}_{noise}$ is very sensitive 
to the choice of $h$ and $c$. The transition from the non-linear to linear increase in $\bar{\lambda}$ in Fig.~\ref{fig:tension_rho}(a) 
corresponds to $\text{Pe} \approx 100$. This value is lower than the theoretical estimate, which is likely due to the presence of frequent 
inter-ABP collisions at $\phi = 0.18$, and enhanced membrane fluctuations facilitated by active particles (see Section~\ref{sec:fluct}). 
Interestingly, a shift between the $\bar{\lambda}$ curves for $\epsilon = 0$ and $\epsilon > 0$ in 
Fig.~\ref{fig:tension_rho}(a) also corresponds to about Pe$\simeq 100$.  

Since membrane tension is affected by the ABP adhesion, we also compute the fraction $\rho$ of particles which are in a direct contact with the 
membrane. Figure \ref{fig:tension_rho}(c) shows that $\rho$ for adhesive ABPs is nearly twice larger than for non-adhesive particles at low $\text{Pe}$. 
As $\text{Pe}$ is increased, $\rho$ rapidly approaches unity for the cases with $\epsilon > 0$ and levels off for $\text{Pe}>100$, while in 
the absence of adhesion, $\rho$ reaches a value of $0.88$ only at $\text{Pe}=400$. Therefore, adhesive interactions make a difference even at large 
$\text{Pe}$. Although the fraction $\rho$ of near-membrane ABPs seem to follow the same trend for $\epsilon= 0$ and $ \epsilon > 0$, the physical 
mechanisms are different. For $\epsilon=0$, an increase in $\text{Pe}$ leads to an increase in the number of ABPs at the membrane 
due to activity-induced accumulation of ABPs at surfaces \cite{Elgeti_WAS_2013,Fily_DSP_2014}. ABPs spend on average more time at the surface 
with increasing $\text{Pe}$, since the escape times decrease with decreasing rotational diffusion, leading to an increase in $\rho$. Furthermore,  
there exists a feedback mechanism between particle accumulation and membrane curvature \cite{vutukuri2020active,iyer2022non}, as the propulsion 
force exerted on the membrane induces a larger local curvature and ABPs accumulate in regions of the large curvature \cite{fily2015dynamics,Iyer_MIPS_2022}.     
For the cases with $\epsilon > 0$, this mechanism is also partially relevant, however, already at low $\text{Pe}$, most of the particles are 
located at the membrane due to adhesive interactions. The fraction $\rho$ at low $\text{Pe}$ for adhesive ABPs in Fig.~\ref{fig:tension_rho}(c)  
does not reach unity because of the strong wrapping of particles by the membrane, whose area is insufficient to all ABPs at $\phi=0.18$.   
As $\text{Pe}$ is increased and ABPs have a sufficient force to detach from the membrane, near-equilibrium 'frozen' structures with strong particle 
wrapping dissolve and the activity-induced accumulation of ABPs results in $\rho$ to approach unity. Note that even though the fraction of ABPs 
at the membrane for $\epsilon > 0$ is larger than that for $\epsilon = 0$, it does not contribute in the same way to membrane tension. For 
$\epsilon=0$, the larger is the fraction $\rho$, the larger is the mean membrane tension $\bar{\lambda}$ due to an increasing swim pressure. 
For $\epsilon > 0$, even though an increase in $\text{Pe}$ leads to an increase in $\bar{\lambda}$ for the same reason, ABP adhesion 
reduces mean membrane tension because of long-ranged adhesive interactions discussed above. Moreover, at low $\text{Pe}$, a number of adhered 
ABPs may temporarily be oriented away from the membrane without detaching from it, which results in a reduction of the total swim pressure.

\subsection{Vesicle shape fluctuations}
\label{sec:fluct}

In the fluctuating regime, we analyse vesicle shape changes by computing the fluctuation spectrum of a membrane cross-section, as 
outlined in Appendix B. Fluctuation spectra of active vesicles at $\phi=0.18$ are presented in Fig.~\ref{fig:fluct} for various $\epsilon$ 
and $\text{Pe}$ values. The fluctuation spectra can be divided into the three regimes with respect to the mode number $l$: (i) low $l \lesssim 10$ where the ABP activity or adhesion dominate, (ii) intermediate $10 \lesssim l \lesssim l_\sigma$ where the competition between the ABP propulsion and adhesion is important, with $l_\sigma = 2\pi R / \sigma \simeq 50$ being a wavelength of the ABP size, and (iii) large $l \gtrsim l_\sigma$ where passive bending rigidity of the membrane dominates.     
At low mode numbers $l \lesssim 10$ and small $\text{Pe}\lesssim 50$ values, a plateau-like region is observed, 
which is more pronounced for large adhesion strengths. This indicates that large-wavelength fluctuations are suppressed in the presence 
of adhesion at low $\text{Pe}$ due to nearly non-dynamic ABP clusters adhered to the membrane. The suppression of large wavelength fluctuations 
has also been observed in cells due to the presence of an underlying cytoskeleton \cite{gov2005red,gov2003cytoskeleton}. Thus, the adhesion 
of particles to the membrane leads to a membrane confinement effect, significantly reducing fluctuations at low $l$ modes. As $\text{Pe}$ 
is increased, the ABPs attain sufficient propulsion force to detach from the membrane, accompanied by the disappearance of the plateau 
region at low $l$. Furthermore, the exponent $\beta$ of fluctuation modes at low $l \in [2,8]$ becomes $\beta < -1$ [see the inset in 
Fig.~\ref{fig:fluct}(a)], which is a clear signature of active membrane fluctuations \cite{takatori2020active,vutukuri2020active}. 
$\beta$ decreases as a function of $\text{Pe}$, demonstrating the enhancement of low-mode fluctuations due to ABP activity. 

A shift in the fluctuation-spectrum curves for different adhesion strengths and intermediate $l$ values at $\text{Pe}=50$ in Fig.~\ref{fig:fluct}(b) 
is likely due to the fact that a number of adhered ABPs can enhance membrane fluctuations by exerting temporary forces in the direction away 
from the membrane without detaching from it. Note that a reduction in tension for $\epsilon > 0$ cannot significantly contribute to this shift 
in fluctuation spectrum, because the effect of membrane tension is expected to be present for $l \lesssim 10-15$ \cite{vutukuri2020active},
while the observed shift extends significantly beyond those $l$ values. Furthermore, for $\text{Pe}=200$ and $\epsilon=2.5k_BT$ in Fig.~\ref{fig:fluct}(b), 
the shift in fluctuation spectrum nearly disappears despite the fact that the mean membrane tension is significantly larger than in the case 
of $\text{Pe}=50$ and $\epsilon=0$ [see Fig.~\ref{fig:tension_rho}(a)]. This suggests that the combination of ABP activity (i.e., applied forces in the direction away 
from the membrane) and adhesion is 
responsible for the shift in fluctuation spectrum for $\text{Pe} \lesssim 100$. 

Another interesting feature in the fluctuation spectra in Fig.~\ref{fig:fluct} for adhesive ABPs with $\epsilon > 0$ is the enhancement of
amplitudes $a_l^2$ at $l\simeq 40-60$ corresponding to the ABP size, since $l_\sigma = 2\pi R/\sigma\simeq 50$. This local enhancement 
in $a_l^2$ represents the wrapping of adhesive ABPs by the membrane, as it is consistently reduced at $\text{Pe}=200$ in comparison to 
$\text{Pe}=50$. Finally, at large $l$, the squared fluctuation amplitudes decay as $l^{-3}$ irrespective of ABP adhesion or $\text{Pe}$, 
corresponding to the bending-dominated regime of membrane fluctuations. 

\begin{figure*}[t!]
    \centering
    \includegraphics{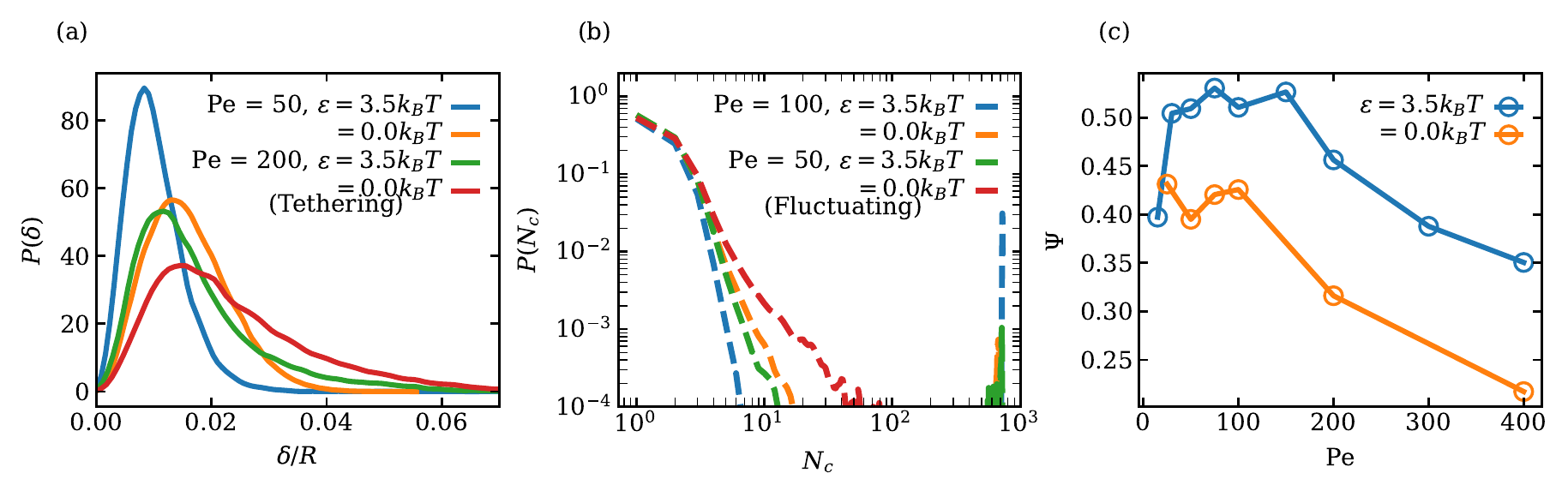}
    \caption{(a) Distributions of fixed-time displacements $\delta$ of single ABPs for different $\epsilon$ and $\text{Pe}$ values in the tethering regime. 
    (b) Distributions of cluster sizes $N_c$ in the fluctuating regime at $\phi=0.18$ for different $\text{Pe}$ and $\epsilon$. 
    (c) Mean cluster asphericity $\Psi$ as a function of $\text{Pe}$ for $\epsilon=0$ and $\epsilon=3.5k_BT$.  }
    \label{fig:clusters}
\end{figure*}

\begin{figure*}[htb!]
    \centering
    \includegraphics{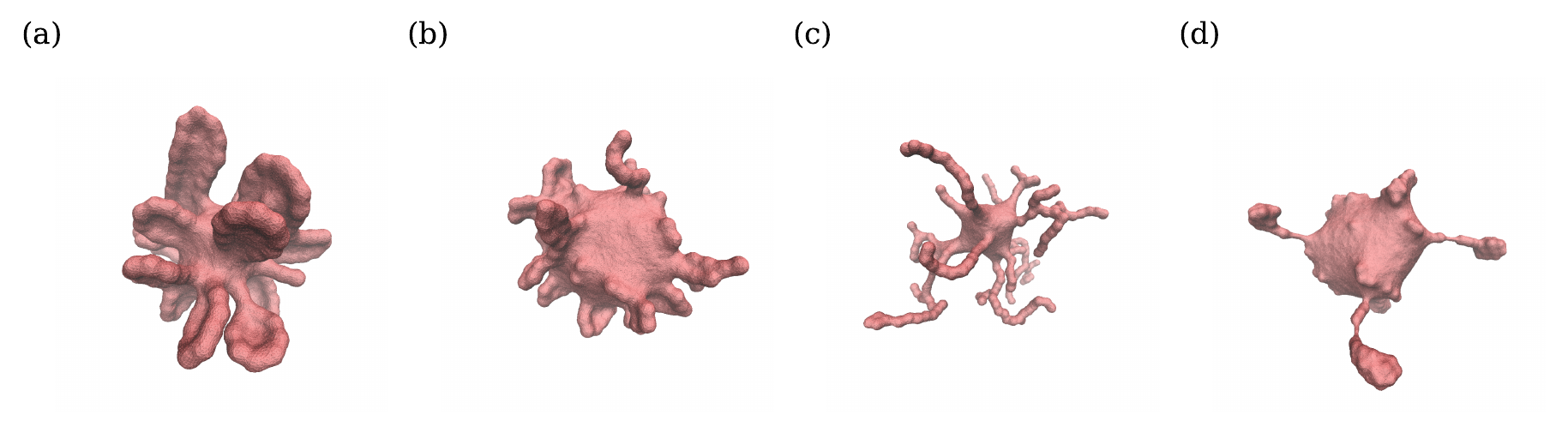}
    \caption{Vesicle shapes for $\phi=0.04$ and $\epsilon=3.5k_BT$ at (a) $\text{Pe}=15$ (see Movie S4), (b) $\text{Pe}=50$, (c) $\text{Pe}=150$ (see Movie S1), and 
    (d) $\text{Pe}=300$. Particle structures change from membrane-wrapped ring-like arrangements to membrane-wrapped (branched) tubular aggregates, 
    as $\text{Pe}$ is increased. A further increase in $\text{Pe}$ leads to the detachment of ABPs from the membrane and their accumulation at 
    the tether end.}
    \label{fig:tethering}
\end{figure*}

\subsection{ABP characteristics}
\label{sec:abp_prop}

Adhesion of ABPs to the membrane must decrease their overall mobility. Figure~\ref{fig:clusters}(a) presents distributions of fixed-time displacements 
$\delta$ of single ABPs for various $\text{Pe}$ and $\epsilon$ values in the tethering regime. As expected, ABP mobility is significantly reduced for 
the cases of $\epsilon > 0$ in comparison to non-adhesive ABPs, and the reduction in particle mobility is more pronounced at low $\text{Pe}$, since 
adhesion interactions dominate over the ABP activity. The mobility of active particles can also be reduced due to the formation of ABP clusters inside
the vesicle. Figure~\ref{fig:clusters}(b) shows distributions of cluster sizes $N_c$ (i.e, the number of ABPs per single cluster) at large $\phi$ 
for various $\text{Pe}$ and $\epsilon$ in the fluctuating regime. In the absence of adhesion ($\epsilon=0$), an increase in $\text{Pe}$ leads to 
an increased accumulation of ABPs at the membrane, such that large clusters are formed through a reduction in the number of small clusters, as 
can be seen through the emergence of a peak at large $N_c$ for $\text{Pe}=100$ in Fig.~\ref{fig:clusters}(b). For $\epsilon > 0$, ABP adhesion to the 
membrane further facilitates the membrane-mediated formation of large particle clusters, as in this case, a peak at large $N_c$ develops already at $\text{Pe}=50$ 
in Fig.~\ref{fig:clusters}(b). As a result, adhesive interactions generally enhance cluster formation in comparison to the case of non-adhesive ABPs. 

We also compute cluster asphericity $\Psi$ (see Appendix C for details) to quantify the effect of ABP adhesion on cluster shapes. Figure~\ref{fig:clusters}(c) 
presents $\Psi$ as a function of $\text{Pe}$, and demonstrates that adhesive interactions cause an increase in the asphericity of 
ABP clusters. Thus, ABP clusters for $\epsilon > 0$ attain shapes, which are further away from a spherical geometry, in agreement with 
the branched string-like arrangements of ABPs in the tethering regime discussed in Section \ref{sec:diag}. For $\epsilon=0$, ABPs primarily cluster 
at the end of tethers as nearly spherical aggregates. Interestingly, $\Psi$ for the case of adhesive ABPs first increases 
and then decreases with increasing $\text{Pe}$. Characteristic vesicle shapes are illustrated in Fig.~\ref{fig:tethering} for different $\text{Pe}$. 
At low $\text{Pe}$, ring-like ABP clusters [Fig.~\ref{fig:tethering}(a)] in the near-equilibrium regime are observed and have the asphericity 
of about $\Psi = 0.4$. With increasing $\text{Pe}$, branched string-like clusters of ABPs within membrane tubes develop with $\Psi > 0.4$, 
see Fig.~\ref{fig:tethering}(c). At large $\text{Pe} \gtrsim 200$, ABP propulsion forces dominate over adhesive interactions, so that the string-like 
structures are destabilized and the ABPs cluster at the tether ends [Fig.~\ref{fig:tethering}(d)] with a reduced cluster asphericity. 
In conclusion, the results in Fig.~\ref{fig:clusters} clearly show that adhesive interactions of ABPs with the membrane strongly alter 
the behavior of individual ABPs and their clusters. 

\section{Summary and conclusions}
\label{sec:concl}

Vesicles enclosing active particles exhibit a variety of dynamic shape deformations, ranging from tethers to prolate and bola-like shapes. Adhesive 
interactions between particles and a vesicle in equilibrium can lead to strong, although static, deformations of the vesicle, such as the formation 
of buds and long tubular structures. In this work, we have combined the effects of particle activity and adhesion to study the deformation and 
properties of vesicles enclosing adhesive ABPs. At low propulsion forces of ABPs, adhesion interactions with the membrane dominate, 
leading to the formation of membrane structures (e.g., buds, tubes) which are similar to those in equilibrium. Furthermore, due to the absence 
of a volume constraint in our simulations, strong membrane deformations with ring-like and sheet-like ABP structures occur for moderate volume 
fractions of ABPs, which are governed by the balance of adhesive interactions and energetic costs for membrane bending. As the propulsion of 
ABPs (or the Peclet number $\text{Pe}$) is increased, the particles are able to detach from the membrane, and the effects of adhesion 
become less dominant. A simple estimation for the detachment force of a single ABP adhered to the membrane based on theoretical arguments 
and simulations yields the adhesion-dominated regime for $\text{Pe} \lesssim 200$. However, ABP-ABP collisions at large enough $\phi$ and 
enhanced membrane fluctuations due to the particle activity further lower the characteristic $\text{Pe} \lesssim 100$ determining the adhesion-dominated 
regime. In the tethering regime, adhesion interactions between the membrane and ABPs significantly reduce the characteristic $\text{Pe}$ for 
tether formation in comparison to non-adhesive ABPs. Furthermore, ABP adhesion favours the formation of long branched tether structures partially 
or fully filled with active particles for low to moderate volume fractions.

At large $\phi$, an increase in $\text{Pe}$ first causes 'melting' of nearly frozen particle structures within the vesicle at low $\text{Pe}$, such that 
the vesicle attains a spherical shape with pronounced membrane fluctuations. A further increase in $\text{Pe}$ results in elongated vesicle shapes or 
bola-like shapes which eventually split into two daughter vesicles. Different from active vesicles with non-adhesive ABPs, for which the fluctuating 
regime is observed at low $\text{Pe}$ across all $\phi$ values, membrane fluctuations in the presence of ABP adhesion take place only at 
$\phi \gtrsim 0.07$ and require some activation energy through a non-zero $\text{Pe}$. The fluctuation spectrum at low $\text{Pe}$ has a plateau 
at low mode numbers because of a 'caging' effect due to the adhered particles. ABP adhesion to the membrane leads to local membrane compression 
with a slightly negative tension due to long-ranged adhesive interactions, so that the mean vesicle tension is lower in the case of adhesive ABPs than 
for non-adhesive particles. With increasing $\text{Pe}$, the mean membrane tension of the vesicle first has a non-linear dependence on $\text{Pe}$ in the adhesion 
dominated regime, followed by a linear increase of the mean tension at large enough $\text{Pe} \gtrsim 100$, in agreement with theoretical 
predictions from the Young-Laplace equation in the case of non-adhesive ABPs \cite{vutukuri2020active,iyer2022non}. Furthermore, the adhesion of ABPs 
to the membrane leads to a reduced particle mobility, but enhances ABP clustering through membrane-mediated interactions. Also, ABP clusters in 
the presence of adhesive interactions have larger cluster asphericities than those for non-adhesive ABPs, mainly due to the formation of branched 
string-like structures of ABPs within membrane tubes in the tethering regime. In conclusion, the presence of adhesive interactions between ABPs and 
the membrane affects not only the phase diagram of active vesicles, but also membrane characteristics (e.g., shape, tension) and ABP properties 
(e.g., mobilite, clustering). Therefore, particle adhesion serves as an additional parameter for the control and tuning of the behavior of active 
vesicles.

\section*{Appendix A: Calculation of membrane tension}
\label{app:tension}
	
Membrane tension is calculated using the virial theorem \cite{tsai1979virial}. The sum over virial contributions from the local area 
constraint is given by 
\begin{equation}
	V_{{\mathrm{av}}} =\sum_{\alpha} \left\langle f^{\mathrm{a},\alpha}_i r^\alpha_i+  f^{\mathrm{a},\alpha}_j r^\alpha_j +  
	f^{\mathrm{a},\alpha}_k r^\alpha_k\right\rangle,
\end{equation}
where $f^{\mathrm{a}}_{i,j,k}$ are forces at the vertices $i$, $j$ and $k$ of a triangle within the membrane triangulation, and  
$\alpha=x$, $y$, or $z$ represents the three coordinates. For elastic bond forces $f^\mathrm{b}$, the virial contribution $V_{\mathrm{b}}$ is
\begin{equation}
	V_{\mathrm{b}}=\sum_{\alpha}\left\langle  f^{\mathrm{b},\alpha}_i r^\alpha_i +f^{\mathrm{b},\alpha}_j r^\alpha_j \right\rangle .
\end{equation}
The total virial contribution from the forces at each vertex is then $V=V_{{\mathrm{av}}}/3  + V_{\mathrm{b}}/2$. The tension of the membrane 
is calculated as a spatial and temporal average of the local stresses as
\begin{equation}
	\lambda =\left \langle \frac{1}{2a_i}( V_i(t) + 2k_{\mathrm{B}}T) \right \rangle_{i,t},
\end{equation}
where the factor two is due to the dimensionality of the membrane, $V_i(t)$ is the virial contribution at vertex $i$ at time $t$, and 
$a_i$ is the area of the dual cell, which is approximated by considering that each neighbouring triangle to the vertex contributes roughly 
one third to the area. The contribution from momentum transfer ($2k_{\mathrm{B}}T$) is approximated by using the equipartition theorem. 
The contributions from the bending energy, volume conservation, and global area conservation are not considered for membrane tension, because 
the bending forces mainly act perpendicular to the tension plane, while the global area and volume constraints have not been used in the simulations.
	
\section*{Appendix B: Membrane shape fluctuations}
\label{app:mem_fluct}

The membrane shape fluctuations are measured by considering 2D sections of the vesicle contour in the x, y, and z directions. The local 
membrane position in these contours is given by $r(\theta_m)$, where $\theta_m = 2\pi m/n$ and $2\pi/n$ is the angle for contour discretisation. 
The fluctuation mode amplitudes $a_l$ are given by the decomposition \cite{pecreaux2004refined,faizi2020fluctuation,faucon1989bending}
\begin{equation}
    a_l = \frac{1}{n}\sum_{m=0}^{n-1} r(\theta_m)\exp\left[\frac{-2\pi i l m}{n}\right].
\end{equation}
The complex modes $a_l$ are calculated using the open-source FFTW \cite{frigo2005design} library, and are averaged over different time frames. 
	
\section*{Appendix C: Asphericity of SPP clusters}
\label{app:cluster_shapes}
	
Shapes of ABP clusters are quantified by their asphericity. The asphericity is calculated from the gyration tensor $G$, which is based 
on the second moments of $N$ particle positions as
\begin{equation}
	G_{xy} \equiv \frac{1}{N} \sum_{i=1}^{N} r^i_x r^i_y ,
\end{equation}
where $\mathbf{r}$ is measured from the center of mass of the $N$-particle system, i.e. $\sum_{i=1}^{N} \mathbf{r}_i = 0 $. Let $\lambda_1$, 
$\lambda_2$, and $\lambda_3$ be the eigenvalues of $G$. Then, the asphericty $\Psi$ is defined as \cite{rudnick1986aspherity}
\begin{equation}
	\Psi = \frac{(\lambda_1-\lambda_2)^2 +(\lambda_2-\lambda_3)^2 + (\lambda_1-\lambda_3)^2}{2(\lambda_1+\lambda_2+\lambda_3)^2}. 
\end{equation}
Values of $\Psi$ range between $0$ and $1$, with $\Psi=0$ for a perfectly spherical shape, $\Psi=1$ for a long thin rod, and $\Psi=0.25$ for a thin plate. 

\section*{Appendix D: Description of movies}

All movies are for an adhesion strength of $\epsilon=3.5k_BT$.

\noindent
\textbf{Movie S1:} Formation of dynamic and highly branched tether structures at $\text{Pe}=150$ and $\phi=0.04$. As ABP 
motion along the tether is limited due to their string-like arrangement, rotational diffusion of the ABPs facilitates tether branching in contrast 
to ABP escape from the tether for $\epsilon=0$.

\noindent
\textbf{Movie S2:} Tether formation at $\text{Pe}=300$ and $\phi=0.009$. ABPs can escape from a tether and join new tethers 
due to their rotational diffusion.

\noindent
\textbf{Movie S3:} Vesicle elongation followed by splitting in the bola regime at $\text{Pe}=200$ and $\phi=0.12$. 

\noindent
\textbf{Movie S4:} Formation of nearly static ring-like structures of ABPs at $\text{Pe}=15$ and $\phi=0.04$.

\section*{Author Contributions}
G.G. and D.A.F. conceived the research project. P.I. performed the simulations and analysed the obtained data. All authors participated 
in the discussions and writing of the manuscript.

\section*{Conflicts of interest}
There are no conflicts to declare.

\section*{Acknowledgements}
We thank Thorsten Auth and Roland G. Winkler for many helpful discussions. The authors gratefully acknowledge the computing time granted through 
JARA on the supercomputer JURECA \cite{jureca} at Forschungszentrum J\"ulich.







%

\end{document}